\documentclass[aps,pra,twocolumn,superscriptaddress,groupedaddress]{revtex4-1}
\usepackage{graphicx,amsmath,amssymb,amsfonts,latexsym,color,dcolumn,bm,mathtools}
\usepackage{xspace}
\newcommand{\beq}{\begin{equation}}
\newcommand{\eeq}{\end{equation}}
\newcommand{\bea}{\begin{eqnarray}}
\newcommand{\eea}{\end{eqnarray}}
\providecommand{\abs}[1]{\left\lvert#1\right\rvert}

\providecommand{\moy}[1]{\langle #1 \rangle}
\providecommand{\expect}[2]{\langle #1 \rangle_{#2}}

\providecommand{\bra}[1]{\langle #1 \rvert}
\providecommand{\ket}[1]{\lvert #1 \rangle}

\providecommand{\tr}[1]{\text{tr}\left[ #1 \right]}

\newcommand{\ketbra}[2]{\lvert #1 \rangle \langle #2\rvert}

\newcommand{\commut}[2]{\ensuremath{\left[ #1 , #2 \right]}\xspace}

\graphicspath{ {/} }

\usepackage{tikz}
\usepgflibrary{arrows}
\usetikzlibrary{decorations}
\usetikzlibrary{decorations.pathmorphing,patterns}
\usetikzlibrary{scopes}
\usetikzlibrary{shapes, matrix}
\begin{document}

\title{Generalized spin squeezing inequalities for particles number
  with  quantum fluctuations} 
\author{I. Saideh}
\email{ibrahim.saideh@u-psud.fr}
\affiliation{Univ Paris Diderot, Sorbonne Paris Cit\'e, MPQ, UMR 7162 CNRS, F-75205 Paris, France}
\affiliation{Institut des Sciences Mol\'eculaires d'Orsay, B\^atiment 350, UMR8214, CNRS-Universit\'e Paris-Sud, Universit\'e Paris-Saclay 91405 Orsay, France}
\author{S. Felicetti}
\affiliation{Univ Paris Diderot, Sorbonne Paris Cit\'e, MPQ, UMR 7162 CNRS, F-75205 Paris, France}
\author{T. Coudreau}
\affiliation{Univ Paris Diderot, Sorbonne Paris Cit\'e, MPQ, UMR 7162 CNRS, F-75205 Paris, France}
\author{P.~Milman}
\affiliation{Univ Paris Diderot, Sorbonne Paris Cit\'e, MPQ, UMR 7162 CNRS, F-75205 Paris, France}
\author{Arne Keller}
\affiliation{Institut des Sciences Mol\'eculaires d'Orsay, B\^atiment 350, UMR8214, CNRS-Universit\'e Paris-Sud, Universit\'e Paris-Saclay 91405 Orsay, France}

\begin{abstract}
Particle number fluctuations, no matter how small, are present in experimental set-ups. One should rigorously take these fluctuations into account, especially, for entanglement detection.
In this context, we  generalize the spin squeezing
inequalities introduced by T\'oth et al. in Ref.\cite{Toth2007}. These new inequalities  are fulfilled by all
separable states even when the number of particle is not constant, and
may present quantum fluctuations.  These inequalities are useful for
detecting  entanglement in many-body systems when the super-selection rule does not
apply,  or when only a subspace of the total  systems Hilbert space is considered. We also define general dichotomic observables for which we obtain a coordinate independent form of the generalized spin squeezing inequalities. We give an example where our generalized coordinate independent spin squeezing inequalities  present a clear advantage over the original ones. 
\end{abstract}

\maketitle

\section{Introduction}
In the quest for quantum computers and quantum simulators, the
ability to create, detect and characterize large scale entanglement in many body systems is one
of the key point that has attracted a lot of interest in the last
decade~\citep{Anders2009,Chen2014,Klempt2014}. 
From a more fundamental perspective, the understanding of the
entanglement properties and their manipulation at the macroscopic
level is also of 
importance to understand the quantum to classical transition~\cite{Milman2007,Aolita2008}.
It is worth noticing that due to the exponential growth of the Hilbert space
dimension with the number of parties $N$,  an exact numerical simulation of such systems with classical
computer is not possible when $N$ becomes of the order of some tens. 
In this context, providing theoretical tools for the experimental
detection of entanglement is a necessity. Typically,  such experiments
involve interacting many
body systems as cold atoms~\cite{meineke2012,Klempt2014}, trapped ions~\cite{haffner2005,Monz2011,bohnet2016quantum} or photons~\cite{Prevedel2009,Iskhakov2012}, and individual
addressing or accessing each body individually is not possible. Accessible
observables consist more often of collectives ones, expressed as the
sum of local observables that in most cases are one body operators.
The spatial component of a collective spin,
sum of local spins, is such an instance of a global observable. If the global spin state is
squeezed, that is, if the fluctuation of one of its component is
sufficiently small compared  to the expectation of the other
components, then it can be shown that the $N$-spin systems is
entangled~\cite{Kitagawa1993,Sorensen1999}. T\'oth and collaborators~\cite{Toth2004,Toth2007,Toth2011, Toth2014} 
have generalized such an approach providing a set of inequalities that are fulfilled for all separable state of
the $N$ spins system and thus are able to detect entanglement when
violated.

The original spin-squeezing inequalities~\cite{Toth2004,
Toth2007,Toth2011} consider  the number of particles $N$ as a
constant. In fact, $N$ may undergo classical and/or quantum
fluctuations. Classical
fluctuations are due  the presence of statistical mixtures of states with
different $N$.
In contrast, quantum fluctuations are given by coherent superposition of states corresponding to
different number of particles. It is often
argued, in the context of
Bose-Einstein 
condensation~\cite{Leggett2001}, that coherent superposition of states corresponding to
different number of particles are not allowed or can not give
observable consequences. The proscription of such coherent
superposition is often justified by an axiomatic superselection
rule (SSR) which should be applied to massive particles but not to massless
ones. Actually, this SSR is a consequence of the lack of a fixed
absolute phase reference~\cite{Bartlett2007}. It has been pointed-out
that such phase reference can be established allowing for instance the coherent
quantum superposition of an atom and a molecule~\cite{Dowling2006}.
Quantum and classical particles number fluctuations have been
considered in the context of quantum metrology~\cite{Hyllus2010}, where the relation of quantum-enhanced
parameter estimation and entanglement is investigated when  the
particles number is only known on average. 
Spin squeezing inequalities for fluctuating $N$ have been
considered in Ref~\cite{Toth2012} but the fluctuations of the total 
number of particles considered in that work were only classical (statistical)
fluctuations and quantum fluctuations were not investigated.

In this work we generalize the original spin squeezing inequality of
Ref.~\cite{Toth2014},  by considering the situation of arbitrary
particle number fluctuations, including quantum and/or classical
ones. 
This generalization is important and necessary in many experimental situations even where the SSR applies.
 Such an interesting example can be found in Ref.~\cite{Klempt2014} where  a system of $N$ spin~1 is considered as
a systems of $N$ spin-$1/2$ by projecting each spin~1 on the subspace spanned
by two magnetic sub-levels. In this subspace, quantum fluctuations
(and not only statistical ones) of the
particle number are expected, and the validity of the original spin
squeezing inequalities
 is not granted.

This paper is organized as follows.
Our main results are presented without proof in section~II. In
section~III, a sketch of the proof of our inequalities is presented,
leaving the technicality to appendices. In section~IV, we consider the
special but important 
case  where the 3 measured observables are dichotomic  observables, that is observables with  only 2
different eigenvalues, as the spin-1/2 component operators. For this
particular case, we show that our inequalities can be put in a coordinate system
independent  form.  In section~V, our inequalities are compared to the original ones~\citep{Toth2007,Toth2014}
in two different cases. In the first example given in section~V, we show that it is incorrect,
 in general, to replace $N$ by its expectation value in spin squeezing inequalities, and that our inequalities should be used instead. In the next example, we study numerically entanglement of $N$ spin~1 state. We find that, when restricting to a subspace, our inequalities show a clear advantage over spin squeezing inequalities~\citep{Toth2011,Toth2014} for $N$ spin~1 particles.

\section{Main results}
We first  recall the original spin inequalities~\cite{Toth2014},
and  how they can be generalized  using 3 collective operators $A_1$, $A_2$ and
$A_3$, instead of the 3 components $J_x$, $J_y$ and $J_z$ of a
collective spin. Finally, we present our new inequalities where
particle number fluctuations are considered.
\subsection{Original spin squeezing inequalities}
For the sake of completeness, we start by recalling the original 
inequalities derived in Ref.~\cite{Toth2014} that are fulfilled by all
separable states:
\begin{subequations}
\label{eq:thothIneq}
\begin{align}
\tilde{\Delta}^2 J_x + \tilde{\Delta}^2 J_y +\tilde{\Delta}^2 J_z
&\geq -Nj^2 \label{eq:1a} \\
(N-1) \left[ \tilde{\Delta}^2J_k + \tilde{\Delta}^2J_l\right] &\geq
\moy{\tilde{J}^2_m} - N(N-1)j^2 \\
\moy{\tilde{J}_l^2 + \tilde{J}_m^2} - N(N-1) j^2 &\leq
(N-1)\tilde{\Delta}^2J_k  \\
\moy{\tilde{J}_x^2 + \tilde{J}_y^2 + \tilde{J}_z^2} &\leq N(N-1)j^2,
\label{eq:1d} 
\end{align}
\end{subequations}
where $l,m$ end $k$ refer to different $x,y$ or $z$ component of the
total spin operator $J_l = \sum_{i=1}^{N} J_l^{(i)}$, sum
of the local spins operators  $J_l^{(i)} = \bigotimes_{i'=1; i\neq i'}^{N} 
\openone^{(i')} \otimes j_k^{(i)}$, where $\openone^{(i')}$ denotes the
identity operator and $j_k^{(i)}$ the $k$ component of the spin
 in the one particle Hilbert space.  The eigenvalues of $(\vec{j}^{(i)})^2$ are $j(j+1)$.
As in  Ref.~\cite{Toth2014}, the notation $\tilde{J}^2_k$ means~:
\beq
\label{eq:tildeDef}
\tilde{J}^2_k = J_k^2 - \sum_{i=1}^{N} (J_k^{(i)})^2 =
\sum_{i\neq j=1}^N J_k^{(i)} J_k^{(j)}
\eeq
and the modified variance is defined as $\tilde{\Delta}^2J_k =
\moy{\tilde{J}^2_k} - \moy{J_k}^2$. 
The 4 inequalities Eqs.~\eqref{eq:thothIneq} can be written in the
following compact
form~\cite{Toth2011}:
\begin{equation}
\label{eq:Mainineq}
(N-1)\sum_{k\notin \mathcal{I}}\tilde{\Delta}^2J_k-\sum_{k\in \mathcal{I}}\moy{\tilde{J}_k^2}\geq -N(N-1)j^2,
\end{equation}
where $\mathcal{I}$ can be any subset of $\left\{ x,y,z\right\}$ (including
the empty set). Each inequality in  Eqs.~\eqref{eq:thothIneq} is
obtained by increasing the number of elements in $\mathcal{I}$ by one,
starting from the empty set.

As it has been shown in  Ref.~\cite{Toth2011}, the vectorial character of the spin is not needed
to obtain  Eq.~\eqref{eq:Mainineq}. Indeed,  a set of
3 collective  observables $A_k$ where $k=1,2,3$ can be used instead, each of them
obtained as a sum of local observable as
$A_k=\sum_{i=1}^{N}A_k^{(i)}$. To be able to derive inequalities as
Eq.~\eqref{eq:Mainineq}, it is only required that
\begin{equation}
\label{upbound}
\sum_{k=1}^3\moy{A_k^{(i)}}^2\leq \alpha^2; \forall i =1,2,\cdots,N,
\end{equation}
which is satisfied by the spin operators $J^{(i)}_k$ with $\alpha=j$.
Then, as it has been shown in Ref.~\cite{Toth2011}, using the Cauchy-Schwartz inequality
\beq
\label{eq:cauchySchwartz}
\moy{A_k}^2\leq N \sum_{i=1}^N\moy{A_k^{(i)}}^2
\eeq
and the
concavity of the variance, we obtain the inequalities
\eqref{eq:Mainineq} where $J_k$ is replaced by $A_k$,  $j$ is replaced
by $\alpha$ and where
$\mathcal{I}$ is any subset of $\left\{
  1,2,3\right\}$, including the empty set. 

\subsection{Fluctuations of particles number}
Note that  Eqs.~\eqref{eq:Mainineq} are derived for a fixed number of
particles $N$. To generalize these equations to include quantum fluctuations of the particle number, we consider that we have $N$ sites $(i=1,2,\cdots,N)$, and
that in each site there is one or zero particles. We define
the local positive operator $\hat{N}^{(i)}$ giving the
number of particle in site $i$; it has only two eigenvalues $0$ or $1$
corresponding to the absence or the presence of a particle. 
Hence, the  collective operator $\hat{N} =
\sum_{i=1}^{N} \hat{N}^{(i)}$ represents the total number of
particles. Our main result is that all separable states fulfil the
following inequalities:
\begin{equation}
\label{eq:OurMainineq}
\left(\moy{\hat{N}}-1\right)\sum_{k\notin \mathcal{I}}\tilde{\Delta}^2A_k-\sum_{k\in
  \mathcal{I}}\moy{\tilde{A}_k^2}\geq -\moy{\hat{N}}\left(\moy{\hat{N}}-1\right)\alpha^2 -\delta,
\end{equation}
where $\delta$ is defined as
\beq
\label{eq:delta}
\delta = \tilde{\Delta}^2A_1 + \tilde{\Delta}^2A_2 +
\tilde{\Delta}^2A_3 + \alpha^2\moy{\hat{N}}   
\eeq
and corresponds to the term added to Eq.~\eqref{eq:Mainineq} when
$N$ is replaced by $\moy{\hat{N}}$. That is, setting $\delta=0$
and replacing $\moy{N}$ by the constant $N$ in Eq.~\eqref{eq:OurMainineq} give us  Eq.~\eqref{eq:Mainineq}.
These inequalities are very
convenient since they are as simple as the original ones. 
Indeed, to test their violation, the same type of measurements realized in the original
inequalities for fixed particle number must be performed.
Eq.~\eqref{eq:OurMainineq} can also be written explicitly, by increasing the
cardinality of $\mathcal{I}$~:
\begin{subequations}
\label{eq:modifIneq}
\begin{align}
\tilde{\Delta}^2A_1+\tilde{\Delta}^2A_2+\tilde{\Delta}^2A_3
&\geq-\alpha^2\moy{\hat{N}}\\
\left(\moy{\hat{N}}-1\right)\left(\tilde{\Delta}^2A_i+\tilde{\Delta}^2A_j\right)-\moy{\tilde{A}_k^2}&
\geq -\alpha^2\moy{\hat{N}}\left(\moy{\hat{N}}-1\right)-\delta  \\
\left(\moy{\hat{N}}-1\right)\tilde{\Delta}^2A_i-\moy{\tilde{A}_j^2}-\moy{\tilde{A}_k^2}
&\geq
-\alpha^2\moy{\hat{N}}\left(\moy{\hat{N}}-1\right)-\delta  \\
\moy{\tilde{A}_1^2}+\moy{\tilde{A}_2^2}+\moy{\tilde{A}_3^2} &\leq
\alpha^2\moy{\hat{N}}\left(\moy{\hat{N}}-1\right)+\delta 
\end{align}
\end{subequations}
We note that the first inequality is exactly the same as
Eq.~\eqref{eq:1a} but with $N$ replaced by $\moy{N}$. 
That is, we can replace $N$ by $\moy{N}$ in
Eq.~\eqref{eq:1a} and it remains a valid equation fulfilled by all
separable states when $N$ is not a constant. We also note  that
Eq.~(\ref{eq:modifIneq}a) can be written as
 $\delta \geq 0$.

 Now, if in a given  experiment $\delta$ is found to be
 positive, then inequalities
 Eqs.~(\ref{eq:modifIneq}b-d) are less tight than the original inequalities
 Eqs.~(\ref{eq:thothIneq}b-d). Hence, a violation of
 Eqs.~(\ref{eq:thothIneq}b-d) can appear without violating
 Eqs.~(\ref{eq:modifIneq}b-d). In other words, The simple substitution of $N$ by 
$\moy{N}$  in the original inequalities Eqs. (1b-d) can give false positive. This is why it is crucial to
 consider the term $\delta$ before to affirming entanglement
 detection. 
In the other case, when $\delta <0$, both inequalities, Eq.~(\ref{eq:thothIneq}a)
 or Eq.~(\ref{eq:modifIneq}a), detect
entanglement, but Eqs.~(\ref{eq:modifIneq}b-d) becomes tighter than  than the original
ones, Eqs.~(\ref{eq:thothIneq}b-d). 
Hence, in this case, the visibility of the violation is higher, which can represent an important advantage from the experimental point of view.

\section{Proof}
We give a sketch of the proof leaving the technical details in
appendix~A and B.
The proof is done in two steps. In the first step, inequalities
fulfilled by all product states $\rho=\bigotimes_{i=1}^N\rho^{(i)}$
are obtained, then in a second step we generalize them to all separable
states using  convexity arguments. 
\subsection{Inequalities for product states}
For the first step, the main objective is to obtain a tighter inequality
than the one obtained in Eq.~\eqref{eq:cauchySchwartz} through the
Cauchy-Schwartz inequality.  
For this, the main idea is to map each local state $\rho^{(i)}$, in the
site $i$,
to a spin~1, or a 3-level state $R^{(i)}$ in an auxiliary  Hilbert space spanned
by $\ket{0^{(i)}}, \ket{1^{(i)}}, \ket{2^{(i)}}$ states as follows~:
\begin{widetext}
\begin{equation}
\label{mappingtoqutrit}
R^{(i)}=n_i\left(\dfrac{{\sigma_0}^{(i)}}{2}+\dfrac{\expect{A^{(i)}_1}{\rho}}{2\eta_i}{\sigma_x}^{(i)}+\dfrac{\expect{A^{(i)}_2}{\rho}}{2\eta_i}{\sigma_y}^{(i)}+\dfrac{\expect{A^{(i)}_3}{\rho}}{2\eta_i}\sigma_z\right)+\left(1-n_i\right)\ket{2^{(i)}}\bra{2^{(i)}}
\end{equation} 
\end{widetext}
where $\sigma_0=\ketbra{0^{(i)}}{0^{(i)}} + \ketbra{1^{(i)}}{1^{(i)}}$
is the projection operator on the qubit subspace spanned by the state $\ket{0^{(i)}}$
and $\ket{1^{(i)}}$ and $\sigma^{(i)}_k (k=x,y,z)$ are the Pauli matrices in the
same subspace. The constant $\eta_i$ is chosen as $\eta_i=
\sqrt{\expect{A^{(i)}_1}{\rho}^2+\expect{A^{(i)}_2}{\rho}^2+\expect{A^{(i)}_3}{\rho}^2}
$, such that the term inside the braket in Eq.~\eqref{mappingtoqutrit}
is a pure state $\ket{\Psi^{(i)}}$.
Therefore the state $R^{(i)}$ can also be written as
\beq
R^{(i)} = n_i\ketbra{\Psi^{(i)}}{\Psi^{(i)}} + (1-n_i)\ket{2^{(i)}}\bra{2^{(i)}},
\eeq
where $n_i$ represents the average occupation number of
the particle in site $i$, that is $n_i = \expect{N^{(i)}}{\rho}$. The
mapping $\rho^{(i)} \rightarrow R^{(i)}$ can be interpreted in the
following way: when there is a  particle  in site $i$, we map its state to a
pure state $\ket{\Psi^{(i)}}$ such  that $\bra{\Psi^{(i)}} \sigma_k
\ket{\Psi^{(i)}} = \frac{1}{\eta_i}\expect{A_k^{(i)}}{\rho}$ and when  there
is no particle we attribute this event to state $\ket{2^{(i)}}$. Averaging over
the occupation of the site $i$ gives us the state $R^{(i)}$.
Using techniques similar to those developed in Ref.~\cite{Saideh2015}
we can prove (see appendix~B) that this mapping is completely
positive and thus $R^{(i)} $ is indeed a state, that is a positive
hermitian operator. 
Note that
$\expect{\sigma_x}{R^{(i)}}=n_i\dfrac{\expect{A^{(i)}_1}{\rho}}{\eta_i}$.
This relation is not exactly what we need. Indeed, if we  sum over all sites $i$, 
$\moy{\sum_{i=1}^{N} \sigma^{(i)}}_{R^{(i)}}$  will
not be simply related to  the expectation of original collective operator
$A_1 = \sum_{i=1}^{N} A_1^{(i)}$, because the pre-factor
$\frac{n_i}{\eta_i}$ depends on the site $i$. It can be shown  (see appendix~A) that 
applying a rotation in the qubit subspace, we can obtain a new state
$R^{(i)'}$ such that $\expect{\sigma_x}{R^{(i)'}} =
\dfrac{\expect{A^{(i)}_1}{\rho}}{\alpha}$, where the factor $\alpha$
does not depends on the site $i$ and is defined as $\alpha^2 =
\text{sup}_{\rho^{(i)}} \left[\sum_{k=1}^{3}
  \expect{A^{(i)}_k}{\rho^{(i)}}^2\right]$ .

Now, we can consider the product state $R'=\bigotimes_{i=1}^NR^{(i)'}$
in the qutrit Hilbert space, and define collective spin
operators: $S_k=\sum_{i=1}^N\sigma_k$, which verify the commutation
relation $\commut{S_k}{S_l}=2i\epsilon_{klm}S_m$. Using the Heisenberg
inequality
$\dfrac{1}{4}\abs{\expect{\commut{A}{B}}{R'}}^2\leq \left(\Delta
  A\right)^2 \left(\Delta B\right)^2$ with $A=S_y$ and $B=S_z$, we can write
\begin{equation}
\label{uncertain}
\abs{\expect{S_x}{R'}}^2=\abs{\sum_{i=1}^N\dfrac{\expect{A^{(i)}_1}{\rho}}{\alpha}}^2=\dfrac{\expect{A_1}{\rho}^2}{\alpha^2}\leq \left(\Delta S_y\right)^2 \left(\Delta S_z\right)^2.
\end{equation}
In that way, we can obtain a tighter inequality for $\expect{A_k}{\rho}^2$
\begin{equation}
\label{eventighterAk}
\expect{A_k}{\rho}^2\leq \expect{N}{\rho}\left(\sum_{i=1}^N\expect{A_k^{(i)}}{\rho}^2\right)+\alpha^2\expect{N}{\rho}\left(\expect{N}{\rho}-\sum_{i=1}^N\expect{N^{(i)}}{\rho}^2\right).
\end{equation}
Using similar techniques (see appendix~A for more details), we obtain
general inequalities fulfilled by all product states:
 \begin{equation}
\label{eq:forproduct}
(\moy{\hat{N}}-1)\sum_{k\notin
  \mathcal{I}}\tilde{\Delta}^2A_k-\sum_{k\in
  \mathcal{I}}\moy{\tilde{A}_k^2}\geq
-\moy{\hat{N}}(\moy{\hat{N}}-1)\alpha^2 -\delta',
\end{equation}
where $\delta'$ is given by
\begin{align}
\label{eq:deltaprimedef}
\delta' &= \alpha^2\moy{\hat{N}} - \sum_{i=1}^{N} \sum_{k\in
  \mathcal{I};k\notin \mathcal{I}} \moy{A_k^{(i)}}^2 \nonumber \\
&+ \moy{\hat{N}}\sum_{i=1}^{N} \left( \sum_{k\in
  \mathcal{I}; k\notin \mathcal{I}} \moy{A_k^{(i)}}^2- \alpha^2 \moy{\hat{N}^{(i)}}^2  \right).
\end{align}
Eq.~\eqref{eq:Mainineq} is recovered when we replace $A_k$ by $J_k$
and $\moy{\hat{N}}$ by $N$, in this case
$\alpha = j$, $\hat{N}^{(i)}=\openone^{(i)}$ and 
$\sum_{k\in
  \mathcal{I}; k\notin \mathcal{I}} \moy{A_k^{(i)}}^2 = j^2$,
therefore $\delta'=0$.  The set of inequalities given by
Eq.~\eqref{eq:forproduct} are valid for any product state. The goal now is
to generalize them for any separable state which can be written as a convex
sum of product states.
\subsection{Generalization to all separable states}
The generalization of inequalities given by Eq.~\eqref{eq:forproduct} to all separable states is not straightforward.
To work around this difficulty, we look for an upper bound $\delta$ to $\delta^\prime€™$, such that when $\delta^\prime$ is replaced by $\delta$ in Eq.~\eqref{eq:forproduct}, the resulting inequalities are easily generalized to all separable states by convexity arguments.

In fact, the last term inside the brackets in
Eq.~\eqref{eq:deltaprimedef} is negative,
therefore  $\delta' \leq \alpha^2\moy{\hat{N}} - \sum_{i=1}^{N} \sum_{k\in
  \mathcal{I};k\notin \mathcal{I}} \moy{A_k^{(i)}}^2$.
In addition, from the definition of the modified variance $\tilde{\Delta}^2$
given by Eq.~\eqref{eq:tildeDef}, it is not difficult to show that for
product states we have
\beq
\tilde{\Delta}^2A_k =
-\sum_{i=1}^{N}\moy{A_k^{(i)}}_{\rho_{\text{prod}}}^2,\quad \rho_{\text{prod}}
\text{ a product state}.
\eeq
We thus obtain the following upper bound for $\delta'$:
\beq
\label{eq:deltaDef}
\delta' \leq \alpha^2\moy{\hat{N}} + \tilde{\Delta}^2A_1 + \tilde{\Delta}^2A_2 +
\tilde{\Delta}^2A_3=\delta,
\eeq
which is the expression for $\delta$, we have given previously in Eq.~\eqref{eq:delta}.

Finally, with this new upper bound, Eq.~\eqref{eq:forproduct} becomes:
 \begin{equation}
\label{eq:newupperbound}
(\moy{\hat{N}}-1)\sum_{k\notin
  \mathcal{I}}\tilde{\Delta}^2A_k-\sum_{k\in
  \mathcal{I}}\moy{\tilde{A}_k^2}\geq
-\moy{\hat{N}}(\moy{\hat{N}}-1)\alpha^2 -\delta.
\end{equation}
It turns out, that these inequalities which are valid for all product
states can be generalized to all separable states by convexity (see
appendix~A).
\section{Coordinate system independent form for dichotomic observables }
Due to Heisenberg uncertainty principle, spin squeezing can not be achieved in all directions. The coordinate independent form of the spin squeezing inequalities~\citep{Toth2007,Toth2014}  allows to detect entanglement without knowing a-priori the direction where the squeezing is maximal.

To illustrate this point, let us recall the squeezing
Hamiltonian~\cite{Kitagawa1993} (see Ref.~\cite{Ma2011} for a review and the references therein): 
\beq
\label{eq:squeezingH}
 H=\chi J_x^2=\chi\sum_{i,j=1}^NJ_x^{(i)}J_x^{(j)}
\eeq
with $\chi$ being some coupling constant.
The above squeezing Hamiltonian has been very well studied, both
theoretically and experimentally, for a system of $N$ spins $1/2$ and
is called one axis twisting
Hamiltonian~\cite{Kitagawa1993,Ma2011}. The state
$\ket{\psi(t)}=e^{-iJ_x^2\theta/2}\bigotimes_{i=1}^N\ket{\frac{1}{2}^{(i)}}$,
where $\theta=2\chi t$, is optimally squeezed along the direction
lying in the $x$-$y$ plane 
making an angle $\phi\approx \frac{1}{2}\arctan\left(N^{-1/3}\right)$
with the $x$-axis, for large $N$~\cite{Kitagawa1993,Ma2011}. This
would suggest that in order to better detect the squeezing in such a
state, using spin squeezing
inequalities~(\ref{eq:1a}-\ref{eq:1d})\cite{Toth2014}, one needs to
measure first and second moments of the rotated spin components
$J_{x'}=\cos(\phi)J_x+\sin(\phi)J_y$,
$J_{y'}=\cos(\phi)J_y-\sin(\phi)J_x$, and $J_{z'}=J_z$. The purpose of
coordinate independent spin squeezing inequalities is to precisely
optimally detect  squeezing without knowing a-priori the optimal
direction $\phi$.

\subsection{Coordinate system independent form of the
spin-squeezing inequalities}  
In this section we recall  the coordinate independent form of spin squeezing inequalities~(\ref{eq:1a}-\ref{eq:1d})~\cite{Toth2014} introduced in~\citep{Toth2007} for spin-$1/2$ and in~\cite{Toth2014} for general spin $j$. First, one needs to define the following matrices~\cite{Toth2014}:
\bea
C_{ij}&=&\frac{1}{2}\expect{J_iJ_j+J_jJ_i}{}\\
\gamma_{ij}&=&C_{ij}-\expect{J_i}{}\expect{J_j}{}\\
Q_{ij}&=&\frac{1}{N}\sum_{k=1}^N\left(\frac{\expect{J_i^{(k)}J_j^{(k)}+J_j^{(k)}J_i^{(k)}}{}}{2}-\frac{j(j+1)\delta_{ij}}{3}\right)
\\
\mathfrak{X}&=&(N-1)\gamma +C-N^2Q
\eea
with $\delta_{ij}$ being the Kronecker delta function. Then, the inequalities inequalities~(\ref{eq:1a}-\ref{eq:1d}) can be written in the following form~\cite{Toth2014}:

\begin{subequations}
\label{eq:thothRIIneq}
\begin{align}
\label{eq:RISS1}
&\tr{\gamma}-Nj \geq 0\\
\label{eq:RISS2}
&(N-1)\tr{\gamma}-N(N-1)j+N^2\frac{j(j+1)}{3}-\lambda_{\text{max}}\left(\mathfrak{X}\right)\geq 0\\
\label{eq:RISS3}
&-\tr{C}+Nj(Nj+1)-N^2\frac{j(j+1)}{3}+\lambda_{\text{min}}\left(\mathfrak{X}\right)\geq 0\\
\label{eq:RISS4}
&-\tr{C}+Nj(Nj+1)\geq 0
\end{align}
\end{subequations}

where $\lambda_{\text{max}}(A)$ and $\lambda_{\text{min}}(A)$ are the maximum and minimum eigenvalues of $A$ respectively.  The key idea for the above inequalities is that $\mathfrak{X}$ is diagonalized via an orthogonal matrix $\mathfrak{O}\in O(3)$, i.e., $\mathfrak{X}=\mathfrak{O}\Lambda\mathfrak{O}^T$ with $\Lambda$ a diagonal matrix. Hence, diagonalizing $\mathfrak{X}$ is equivalent to applying the following transformation: $\vec{J'}=\mathfrak{O}^T\mathfrak{J}$, $C'=\mathfrak{O}^T C\mathfrak{O}$, $\gamma'=\mathfrak{O}^T \gamma\mathfrak{O}$, and $Q'=\mathfrak{O}^T Q\mathfrak{O}$. Finally, we have $\tr{C'}=\tr{C}$ and $\tr{\gamma'}=\tr{\gamma}$ which represents the invariance of the quantities $\expect{J_x^2}{}+\expect{J_y^2}{}+\expect{J_z^2}{}$ and $\left(\Delta J_x\right)^2+\left(\Delta J_y\right)^2+\left(\Delta J_z\right)^2$, respectively, under rotations. Comparing the inequalities~\eqref{eq:thothRIIneq} with~\eqref{eq:thothIneq}, one can show that there exists a direction for which some of the inequalities~\eqref{eq:thothIneq} is violated iff the corresponding inequalities~\eqref{eq:thothRIIneq} are violated~\cite{Toth2007,Toth2014}.

A natural question arises whether we could define our generalized
inequalities~\eqref{eq:modifIneq} in a coordinate independent manner
to simplify the task of finding operators $A_i$ such that entanglement
is detected. It turns out that we are able to define a coordinate
independent version inequalities of~\eqref{eq:modifIneq} for a general
class of dichotomic operators which we define now.

\subsection{Dichotomic observables}
A very important and popular choice of the operators 
$A_i$ for the inequalities~\eqref{eq:modifIneq} are the dichotomic
observables or spin-$1/2$ like operators. 
In this case, the Hilbert space of the single particle states is
usually restricted to a 2-dimensional subspace 
of a two levels system. This restriction to a bi-dimensional subspace
has been performed in Ref.~\citep{Klempt2014}
where a system of $N$ spin-$1$ is considered as $N$ spin-$1/2$
particles. The appeal of this choice is due to the fact that most of
the entanglement criteria were originally  derived for spin~$1/2$
systems.  Notorious examples are  CHSH inequalities~\cite{CHSH} for the
non-locality of a two spins-$1/2$ state and spin squeezing
inequalities for $N$ spin-$1/2$~\cite{Toth2007}.

Specifically, for each particle $(i)$, consider only two magnetic
levels states $\ket{m_0^{(i)}}$ and $\ket{m_1^{(i)}}$, among all the
eigenstates $\ket{m^{(i)}}  (\abs{m^{(i)}} \le j)$ of $j_z^{(i)}$. In the subspace spanned by these two states, we can define the Pauli operators as:
\bea
\label{1}
&\sigma_x^{(i)}&=\ketbra{m_0^{(i)}}{m_1^{(i)}}+\ketbra{m_1^{(i)}}{m_0^{(i)}}\\
&\sigma_y^{(i)}&=-i\left(\ketbra{m_0^{(i)}}{m_1^{(i)}}-\ketbra{m_1^{(i)}}{m_0^{(i)}}\right)\\
&\sigma_z^{(i)}&=\ketbra{m_0^{(i)}}{m_0^{(i)}}-\ketbra{m_1^{(i)}}{m_1^{(i)}}
\eea
 and let us call $N^{(i)}$ the projector into this subspace spanned by
 $\ket{m_0^{(i)}},\ket{m_1^{(i)}}$, that is:
 \beq
 \label{2}
 N^{(i)}=\ketbra{m_0^{(i)}}{m_0^{(i)}}+\ketbra{m_1^{(i)}}{m_1^{(i)}}.
 \eeq
 An elementary calculation shows that for any state $\rho^{(i)}$ acting on the single particle Hilbert space, we have:
 \beq
 \label{a}
 \expect{\sigma_x^{(i)}}{\rho^{(i)}}^2+\expect{\sigma_y^{(i)}}{\rho^{(i)}}^2+\expect{\sigma_z^{(i)}}{\rho^{(i)}}^2\leq \expect{N^{(i)}}{\rho^{(i)}}^2.
 \eeq
 Since $N^{(i)}$ is a projector, it is positive and has two eigenvalues $0$ and $1$. Hence, from Eq.~\eqref{a}, if we choose the state $\rho^{(i)}$ to be a pure state in the subspace $\ket{m_0^{(i)}},\ket{m_1^{(i)}}$, we find that
 \beq
 \label{b}
\alpha^2 =
\text{sup}_{\rho^{(i)}} \left[\sum_{k=1}^{3} \expect{\sigma^{(i)}_k}{\rho^{(i)}}^2\right]=1.
 \eeq
We now define the collective operators $A_i$ to be:
\bea
 \label{eq:choice}
 A_i=\frac{1}{2}\sum_{k=1}^N\sigma_i^{(k)}.
 \eea
These three collectives observables $A_i$ fulfill all the requirements to write
the generalized spin squeezing inequalities~\eqref{eq:modifIneq}.
 
The class of dichtotomic observables~\eqref{eq:choice} can be extended
in slightly more general manner. Instead of the two states
$\ket{m_0^{(i)}}, \ket{m_1^{(i)}}$, we can consider two orthogonal projectors $P_0^{(i)}$ and $P_1^{(i)}$ such that
 \beq
 \text{rank}\left(P_0^{(i)}\right)=\text{rank}\left(P_1^{(i)}\right)=r.
\eeq 
  Let us define $\mathcal{H}_0^{(i)}$, $\mathcal{H}_1^{(i)}$ to be the range of $P_0^{(i)}$ and $P_1^{(i)}$ respectively. Let $S^{(i)}$ be a linear map from $\mathcal{H}_1^{(i)}$ to $\mathcal{H}_0^{(i)}$ with singular values equal to $1$, i.e. it can be written as
\beq
S^{(i)}={U^{(i)}}^{\dagger}\openone^{(r)}{V^{(i)}},
\eeq
with 
${U^{(i)}}^{\dagger}U^{(i)}=\openone^{(r)}$ and
${V^{(i)}}^{\dagger}V^{(i)}=\openone^{(r)}$.

Finally, let us define
\beq
S_-^{(i)}=P_0^{(i)}S^{(i)}P_1^{(i)} \,,\,S_+^{(i)}={S_-^{(i)}}^{\dagger}.
\eeq
Now, we can generalize the operators $A_i$ defined in Eq.\eqref{eq:choice} by defining the Pauli-like operators for each particle as:
\bea
\label{eq:GPM}
&\sigma_z^{(i)}&=P_0^{(i)}-P_1^{(i)}\\
&\sigma_x^{(i)}&=S_-^{(i)}+S_+^{(i)}\\
&\sigma_y^{(i)}&=-i\left(S_-^{(i)}-S_+^{(i)}\right)\\
&N^{(i)}&=P_0^{(i)}+P_1^{(i)}.
\eea
With the above definitions, Eqs.~(\ref{a},~\ref{b}) are valid. Moreover, the commutation relations $\commut{\sigma_i^{(l)}}{\sigma_j^{()}}=2i\epsilon_{ijk}\sigma_k^{(l)}$ still hold. Consequently, the operators defined in Eqs.\eqref{eq:GPM} are the generators of SU(2) in the subspace $\mathcal{H}_0^{(i)}\oplus\mathcal{H}_1^{(i)}$, i.e. any rotation can be applied via the unitary $e^{i\vec{\hat{A}}.\vec{n}\theta}$, for some normalized vector~$\vec{n}$ and real $\theta$. In addition, since $\commut{\hat{N}}{A_i}=0$, $\expect{\hat{N}}{}$ is invariant under such rotations and behaves simply as a scalar as in the usual spin squeezing inequalities. Hence, the additional quantity $\delta$ in our inequalities~\eqref{eq:modifIneq}, that takes into account the fluctuations in particle number in the subspace of interest, is also invariant under $SU(2)$ transformations. Which allows us to follow the same steps as in~\cite{Toth2007,Toth2014} to define coordinate system independent inequalities:
 \begin{subequations}
\label{eq:RIIneq}
\begin{align}
\label{eq:GRISS1}
\delta=&\tr{\gamma}-{\expect{\hat{N}}{}}/{2} \geq 0\\
\label{eq:GRISS2}
\delta+&\left(\expect{\hat{N}}{}-1\right)\tr{\gamma}-\expect{\hat{N}}{}\left(\expect{\hat{N}}{}-2\right)/2-\lambda_{\text{max}}\left(\mathfrak{X}\right)\geq 0\\
\label{eq:GRISS3}
\delta-&\tr{C}-{\expect{\hat{N}}{}}/{2}+\lambda_{\text{min}}\left(\mathfrak{X}\right)\geq
0\\
\label{eq:GRISS4}
\delta-&\tr{C}+\expect{\hat{N}}{}\left(\expect{\hat{N}}{}+2\right)/4\geq 0,
\end{align}
\end{subequations}
where we have defined:
\bea
C_{ij}&=&\frac{1}{2}\expect{A_iA_j+A_jA_i}{},\nonumber\\
\gamma_{ij}&=&C_{ij}-\expect{A_i}{}\expect{A_j}{}\\
\mathfrak{X}&=&(N-1)\gamma +C.\nonumber
\eea
As expected, comparing our inequalities~\eqref{eq:RIIneq} with the coordinate independent spin squeezing inequalities for $j=1/2$ in Refs.~\cite{Toth2007,Toth2014}, they are quite similar except for replacing $N$ with $\expect{\hat{N}}{}$ and the additional term $\delta$. Simply replacing $N$ with $\expect{\hat{N}}{}$ is not enough to obtain our inequalities as we will show later.

\begin{figure}[]
\includegraphics[width=0.5\textwidth]{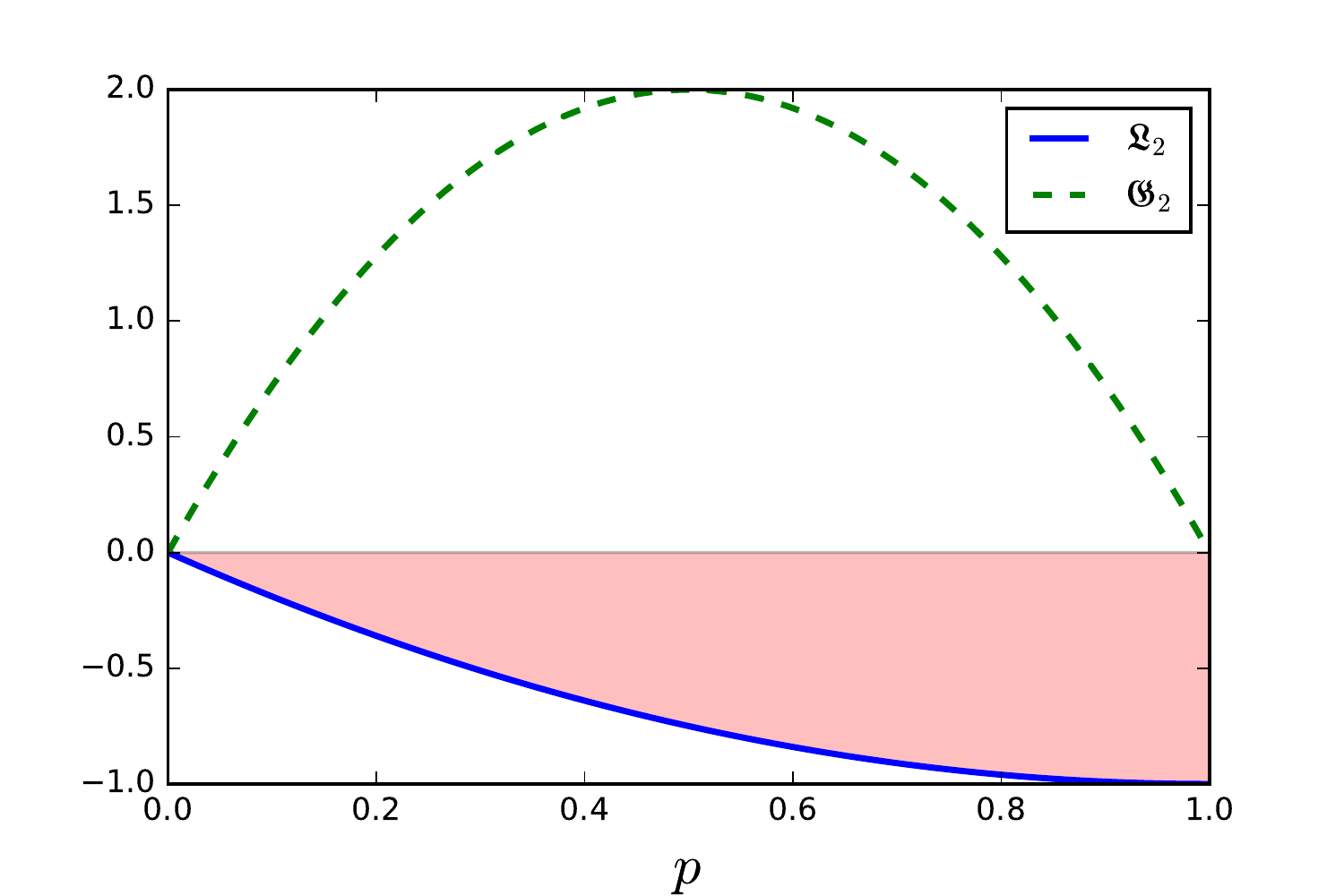}
\caption{(Color online) $\mathfrak{L}$~\eqref{eq:L}, in solid blue line, and  $\mathfrak{G}$~\eqref{eq:G}, in dashed green line, calculated in the state $\rho(p)$~\eqref{eq:mixedrho} as a function of $p$. The highlighted area represents the instances of $p$ for which the inequality $\mathfrak{L}(p)\geq 0$ is violated.}
\label{fig:3}
\end{figure}

\section{Examples}
In this section we compare, for two specific cases, the standard spin
squeezing inequalities where $N$ is replaced by 
its expectation $\moy{N}$ with our new inequalities.
The first case illustrates the importance of the term $\delta$ in our
inequalities.  Indeed, we exhibit a separable mixed state that
violates the original inequality, showing that the simple replacement
of $N$ by $\moy{N}$ can lead to false positive.
In the second example, we study the detection of entanglement
generated by the one axis twisting Hamiltonian~\eqref{eq:squeezingH}
for $N=5$ spin~1 system.  We find that, when restricting to a
subspace, our inequalities~\eqref{eq:RIIneq} show a clear advantage
over spin squeezing inequalities~\eqref{eq:thothIneq}. The latter show
no violation at all, whereas, one of the
inequalities~\eqref{eq:RIIneq} is violated indicating entanglement
almost for all times of the evolution of the $N=5$ spin~1 system.
\subsection{Example I}
We have shown, that through our special choice of operators $A_i$
given by Eq.~\eqref{eq:choice}, our inequalities~\eqref{eq:modifIneq} and ~\eqref{eq:RIIneq} can be obtained from spin squeezing inequalities for $j=1/2$~\cite{Toth2007,Toth2014} by replacing $N$ with $\expect{\hat{N}}{}$ and adding $\delta$. In the following, we give a simple example to highlight the importance of the additional term $\delta$. Let us consider the following separable mixed states for $N=2$ spin $j=1$:
\beq
\label{eq:mixedrho}
\rho(p)=p\rho'+(1-p)\ketbra{0^{(1)}}{0^{(1)}}\otimes\ketbra{0^{(2)}}{0^{(2)}} \quad 0\leq p\leq 1
\eeq
where:
\begin{align}
\rho'&=\frac{1}{2}\left(
\ketbra{-1^{(1)}}{-1^{(1)}}\otimes\ketbra{-1^{(2)}}{-1^{(2)}}
\right. \nonumber \\
&+\left.\ketbra{1^{(1)}}{1^{(1)}}\otimes\ketbra{1^{(2)}}{1^{(2)}}\right)
\end{align}
This state is clearly separable for any value of $0\leq p \leq 1$. Now
let us consider the inequality given by Eq.~(\ref{eq:thothIneq}b)
and let us replace $N$ by $\expect{\hat{N}}{}$ where $\hat{N}$ is defined in Eq.~\eqref{2} for the subspace $\ket{-1^{(i)}},\ket{1^{(i)}}$. Then, it takes the following form: $\mathfrak{L}(p)\geq 0$, where
\begin{align}
\label{eq:L}
\mathfrak{L}(p)=&\left(\expect{\hat{N}}{\rho(p)}-1\right) \left(\tilde{\Delta}^2A_x+\tilde{\Delta}^2A_y\right)
-\expect{\tilde{A}_z^2}{\rho(p)} \nonumber \\
&+\frac{1}{4}\expect{\hat{N}}{\rho(p)}\left(\expect{\hat{N}}{\rho(p)}-1\right)
\end{align}

Next, let us consider the correct form, i.e. Eq.~(\ref{eq:modifIneq}b): $\mathfrak{G}(p)\geq 0$, where 
\beq
\label{eq:G}
\mathfrak{G}(p)=\mathfrak{L}(p)+\delta(p)
\eeq
In figure Fig.~\ref{fig:3}, we plot both quantities $\mathfrak{G}(p)$
and $\mathfrak{L}(p)$ as a function of $p$. Inequality
$\mathfrak{L}(p)\geq 0$ is violated for all $p$, but it is completely
wrong to infer that the state is entangled. In
contrast, our inequality
$\mathfrak{G}(p)=\mathfrak{L}(p)+\delta(p)\geq 0$ is not violated, as
expected. This example shows clearly the importance of the additional
term $\delta$ when the number of particles is not  constant.

\subsection{Example II}

\begin{figure}[t!]
\includegraphics[width=0.5\textwidth]{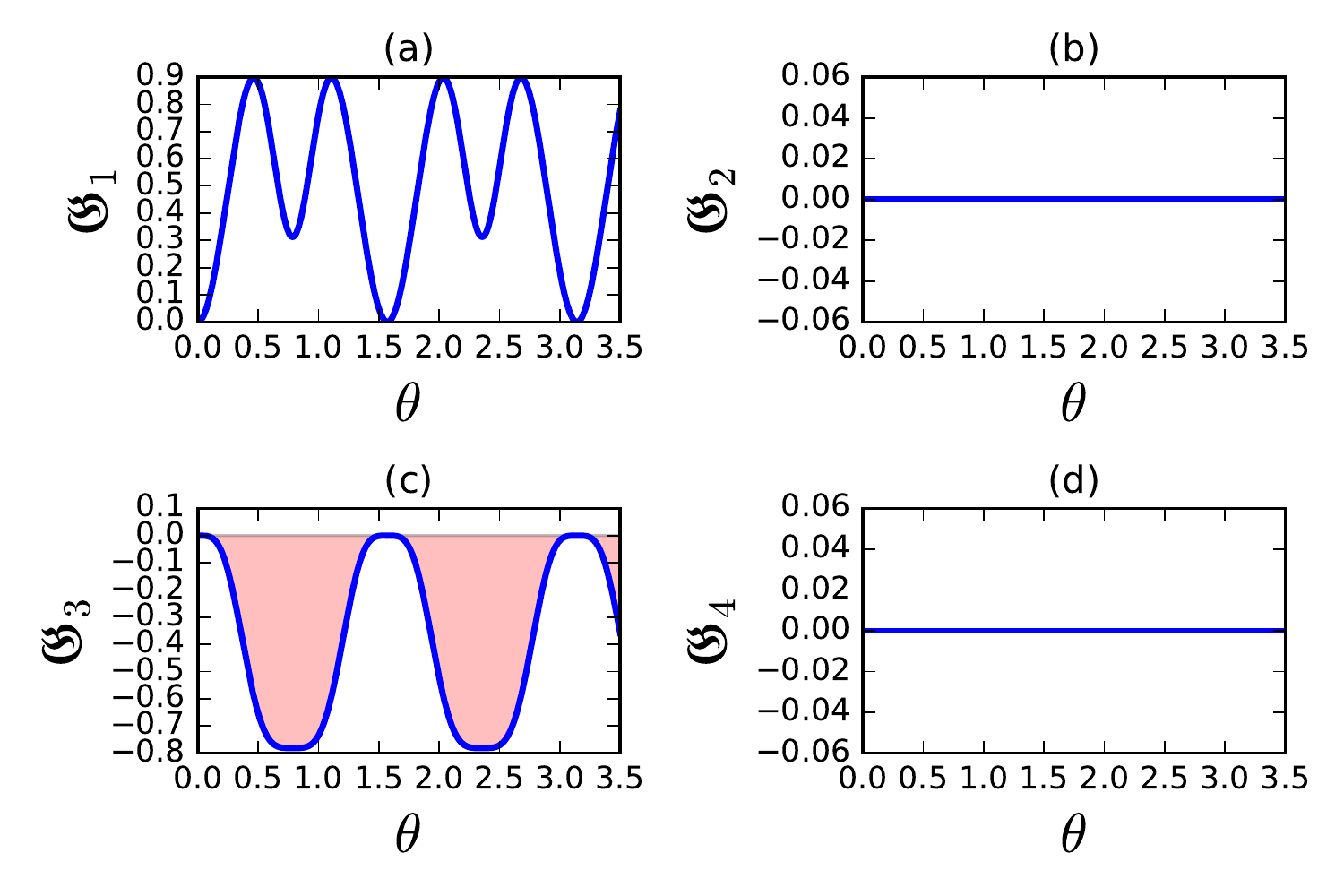}
\caption{(Color online) (a) to (d): left hand side of inequalities $\mathfrak{G}_1$~\eqref{eq:GRISS1} to $\mathfrak{G}_4$~\eqref{eq:GRISS4}, respectively, calculated in the state $\ket{\psi(\theta)}$, defined in Eq.~\eqref{eq:psi}, as a function of $\theta$ for $N=5$ spin-$1$ particles. Highlighted region shows the instances of $\theta$ for which inequality~\eqref{eq:GRISS3} is violated.}
\label{fig:2}
\end{figure}

As an illustrative example, we consider a system of $N$ spin $j=1$ initialized in the product state $\ket{\psi_0}=\bigotimes_{i=1}^N\ket{0^{(i)}}$.
Now let us calculate the left hand side of the inequalities~\eqref{eq:thothRIIneq} for the state
\beq
\label{eq:psi}
\ket{\psi(\theta)}=e^{-iJ_x^2\theta/2}\bigotimes_{i=1}^N\ket{0^{(i)}}
\eeq
 of $N$ spins $j=1$. Let us call $\mathfrak{F}_1(\theta)$, $\mathfrak{F}_2(\theta)$, $\mathfrak{F}_3(\theta)$, and $\mathfrak{F}_4(\theta)$ to be the left hand side of inequalities~(\ref{eq:RISS1}-\ref{eq:RISS4}) respectively. Numerical calculations show that these quantities are constant and positive. More precisely, one can verify that $\mathfrak{F}_1(\theta)=N$ and $\mathfrak{F}_i(\theta)=N(N-1)$ for $i=2,3,4$, thus, spin squeezing inequalities~\eqref{eq:thothRIIneq} fail to detect entanglement in the state  $\ket{\psi(\theta)}$. The constancy of the quantities  $\mathfrak{F}_l(\theta)$, with $l=1,\cdots,4$, is due to the choice of the initial state   $\ket{\psi(0)}=\bigotimes_{i=1}^N\ket{0^{(i)}}$ and the fact we have chosen $j=1$. Non trivial evolution of the quantities $\mathfrak{F}_l(\theta)$  will occur for different initial states and different integer spin values $j=2,3,\cdots$.

However, if we choose different observables than the collective spin components, our generalized inequalities~\eqref{eq:modifIneq} can be violated inferring entanglement of the state $\ket{\psi(\theta)}$ for some $\theta$. In particular, we will define dichotomic observables in the subspace $\ket{-1^{(i)}},\ket{1^{(i)}}$, by setting $\ket{m_0^{(i)}}=\ket{-1^{(i)}}$ and $\ket{m_1^{(i)}}=\ket{1^{(i)}}$ in Eqs.(\ref{eq:choice},~\ref{1},~\ref{2}), so that the $N$ spin-$1$ particles can be seen as $\expect{\hat{N}}{}$ spin-$1/2$ particles.

Now, let us call $\mathfrak{G}_i\left(\theta\right)\,:i=1,2,3,4$ to be the left hand side of inequalities~(\ref{eq:GRISS1}-\ref{eq:GRISS4}), respectively, calculated for the state $\ket{\psi(\theta)}$~\eqref{eq:psi}. In Fig.~\ref{fig:2}, we plot $\mathfrak{G}_i\left(\theta\right)$ for $N=5$, and we can see that $\mathfrak{G}_3\left(\theta\right)$ violates the inequality~\eqref{eq:GRISS3}, Fig.~\ref{fig:2}(c). Consequently, we show that the state $\ket{\psi(\theta)}$ is entangled, at least, when inequality~\eqref{eq:GRISS3} is violated.

\section{Conclusion}
We have generalized the spin squeezing inequalities in order to
consider quantum  fluctuations of the number of particles $N$. Our generalized
inequalities can be obtained from the original ones  by replacing  $N$
with its expectation value $\moy{N}$ and by adding a new  term
$\delta$ which is not more difficult to measure than the other terms
forming the original inequalities.
In the case where the measured observables are dichotomic, we have
shown that we can define coordinates independent spin squeezing
inequalities in the same way it had been defined previously for the
original inequalities. 
The non conservation of the number of particles allows more flexibility
in  the set of observables to be used to test the inequalities. We have
presented an example where such flexibility allows for the detection of an
entangled state which was not detected by the original inequalities. 
We also warn that using the original inequalities, in a context where
the number of particles $N$ fluctuates, by replacing $N$ by its
expectation value $\moy{N}$ can result in a violation for separable
states, hence giving false positive.

\bibliography{spinSqueezing}

\appendix
\section{Proof of Eq.~\eqref{eq:OurMainineq}}\label{Ap:proof}
In this appendix, we present in detail the different steps to derive our main inequality~\eqref{eq:OurMainineq}. As mentioned in the main text, we proceed to the proof in two steps. Firstly, we start by proving the inequality~\eqref{eq:OurMainineq} for product states. Next, we generalize the inequality for mixed state by convexity argument. 
Before proceeding to the two parts of the proof, let us rewrite the inequality~\eqref{eq:OurMainineq} and its different ingredients:
\begin{equation}
\left(\moy{\hat{N}}-1\right)\sum_{k\notin \mathcal{I}}\tilde{\Delta}^2A_k-\sum_{k\in
  \mathcal{I}}\moy{\tilde{A}_k^2}\geq -\moy{\hat{N}}\left(\moy{\hat{N}}-1\right)\alpha^2 -\delta,
\end{equation}
with $\alpha^2 =
\text{sup}_{\rho^{(i)}} \left[\sum_{k=1}^{3} \expect{A^{(i)}_k}{\rho^{(i)}}^2\right]$ and $\hat{N}=\sum_{i=1}^NN^{(i)}$ represents the particle number operator as explained in the main text. We choose the operator $N^{(i)}$ to be positive and to verify the following inequality:
\beq
\label{n}
\frac{\sum_{k=1}^{3} \expect{A^{(i)}_k}{\rho^{(i)}}^2}{\alpha^2}\leq \expect{N^{(i)}}{\rho^{(i)}}^2\leq 1
\eeq
 for any state $\rho^{(i)}$ acting on the single particle Hilbert space. One can always find a positive operator $N^{(i)}$ such that Eq.~\eqref{n} is verified, since one can always choose $N^{(i)}$ to be the identity in the single particle Hilbert space. Finally, we recall  the expression for $\delta$~\eqref{eq:deltaDef}:
\beq
\delta= \alpha^2\moy{\hat{N}} + \tilde{\Delta}^2A_1 + \tilde{\Delta}^2A_2 +
\tilde{\Delta}^2A_3
\eeq
\subsection{Proof of Eq.~\eqref{eq:OurMainineq} for product states} 
As we have outlined in the main text, our main improvement comes from
deriving a new bound for $\expect{A_i}{}^2$ better than the standard
one 
\beq
\label{eq:obvousBound}
\expect{A_i}{}^2\leq N\sum_{k=1}^N\expect{A_i^{(k)}}{}^2.
\eeq
 The
previous inequality can be obtained directly from Cauchy-Schwartz
inequality. However, it can also be obtained in a different way using
the  Heisenberg uncertainty inequality as follows. First, for the sake
of illustration,
consider that  $A_i=S_x=\sum_{i=1}^{N}\sigma_x^{(i)}$ and let
$\ket{\psi}$, the  product state
$\ket{\psi}=\bigotimes_{i=1}^N\ket{\psi^{(i)}}$ of $N$
spin-$1/2$. Starting  from the Heisenberg uncertainty inequality:
\beq
{\expect{S_x}{}^2}\leq \left(\Delta S_y\right)^2\left(\Delta S_z\right)^2,
\eeq
we can  apply a rotation $U^{(i)}=e^{i\frac{\sigma_x^{(i)}\theta_i}{2}}$ to each spin such that 
\[%
\expect{\sigma_y^{(i)}}{U^{(i)}\ket{\psi^{(i)}}}=0
,\expect{\sigma_x^{(i)}}{U^{(i)}\ket{\psi^{(i)}}}=\expect{\sigma_x^{(i)}}{\ket{\psi^{(i)}}}.
\]
Since $\ket{\psi}$ is a pure product state, we have:
\[%
\expect{\sigma_z^{(i)}}{U^{(i)}\ket{\psi^{(i)}}}^2=1-\expect{\sigma_x^{(i)}}{\ket{\psi^{(i)}}}^2.
\]
Then, a straightforward calculation, in the rotated state, would yield $\left(\Delta S_y\right)^2=N$ and $\left(\Delta S_z\right)^2=\sum_{i=1}^N\expect{\sigma_x^{(i)}}{\ket{\psi^{(i)}}}^2$, hence:
\beq
\label{ineq:Sx}
{\expect{S_x}{\ket{\psi}}^2} \leq N \sum_{i=1}^N\expect{\sigma_x^{(i)}}{\ket{\psi^{(i)}}}^2
\eeq
which is the same inequality than Eq.~\eqref{eq:obvousBound}.

It is the above reasoning that motivates the mapping of the original product state $\rho=\bigotimes\rho^{(i)}$ to the the product state of $N$ spin~1, $R=\bigotimes_{i=1}^NR^{(i)}$, where:
\beq
\label{eq:Ri}
R^{(i)} = n_i\ketbra{\Psi^{(i)}}{\Psi^{(i)}} + (1-n_i)\ketbra{2^{(i)}}{2^{(i)}},
\eeq
 $n_i=\expect{N^{(i)}}{\rho^{(i)}}$, and $\ket{\Psi^{(i)}}$ is a pure state defined, in the subspace spanned by $\ket{0^{(i)}}, \ket{1^{(i)}}$, as:
 \beq
\ketbra{\Psi^{(i)}}{\Psi^{(i)}}= \dfrac{{\sigma_0}^{(i)}}{2}+\dfrac{\expect{A^{(i)}_1}{\rho}}{2\eta_i}{\sigma_x}^{(i)}+\dfrac{\expect{A^{(i)}_2}{\rho}}{2\eta_i}{\sigma_y}^{(i)}+\dfrac{\expect{A^{(i)}_3}{\rho}}{2\eta_i}\sigma_z
 \eeq

where $\sigma_0=\ketbra{0^{(i)}}{0^{(i)}} + \ketbra{1^{(i)}}{1^{(i)}}$ and $\sigma^{(i)}_k (k=x,y,z)$ are the Pauli matrices in the
same subspace. The constant $\eta_i$ is chosen as $\eta_i=
\sqrt{\expect{A^{(i)}_1}{\rho}^2+\expect{A^{(i)}_2}{\rho}^2+\expect{A^{(i)}_3}{\rho}^2}
$, to ensure the purity of the state $\ket{\Psi^{(i)}}$. 
\subsubsection{Inequality for $\expect{A_i}{}^2$}
Following the same reasoning as above, We first apply the following unitary $\ketbra{2^{(i)}}{2^{(i)}}+e^{-i\theta_i{\sigma}_x^{(i)}/2}$ to the state $R^{(i)}$ Eq.~\eqref{eq:Ri}.
After applying the unitary, we get the following state:
\begin{equation}
r^{(i)}=n_i\ketbra{\Phi^{(i)}}{\Phi^{(i)}}+\left(1-n_i\right)\ketbra{2^{(i)}}{2^{(i)}},
\end{equation}
where $\ket{\Phi^{(i)}}=e^{-i\theta_i{\sigma}_x^{(i)}/2}\ket{\Psi^{(i)}}$. We choose $\theta_i$ such that:
\begin{align}
\bra{\Phi^{(i)}}{\sigma}_y^{(i)}\ket{\Phi^{(i)}} &\equiv\cos{\theta_i}\bra{\Psi^{(i)}}{\sigma}_y^{(i)}\ket{\Psi^{(i)}}\nonumber
\\
&-\sin{\theta_i}\bra{\Psi^{(i)}}{\sigma}_z^{(i)}\ket{\Psi^{(i)}}=0
\end{align}
which can be achieved  with the choice \[\cos\theta_i=\dfrac{\bra{\Psi^{(i)}}{\sigma}_z^{(i)}\ket{\Psi^{(i)}}}{\sqrt{\bra{\Psi^{(i)}}{\sigma}_z^{(i)}\ket{\Psi^{(i)}}^2+\bra{\Psi^{(i)}}{\sigma}_y^{(i)}\ket{\Psi^{(i)}}^2}},\]

\[\sin{\theta_i}=\dfrac{\bra{\Psi^{(i)}}{\sigma}_y^{(i)}\ket{\Psi^{(i)}}}{\sqrt{\bra{\Psi^{(i)}}{\sigma}_z^{(i)}\ket{\Psi^{(i)}}^2+\bra{\Psi^{(i)}}{\sigma}_y^{(i)}\ket{\Psi^{(i)}}^2}}\]
Since $\commut{{\sigma}_x^{(i)}}{e^{-i\theta_i{\sigma}_x^{(i)}/2}}=0$, we have:
\begin{equation}
\bra{\Phi^{(i)}}{\sigma}_x^{(i)}\ket{\Phi^{(i)}}=\bra{\Psi^{(i)}}{\sigma}_x^{(i)}\ket{\Psi^{(i)}}.
\end{equation}

Now, because $\dfrac{\eta_i}{\alpha n_i}\leq 1$~\eqref{n}, there exists an angle $\xi_i$ such that:
\begin{equation}
\bra{\Phi^{(i)}}e^{i\xi_i{\sigma}_y^{(i)}/2}{\sigma}_x^{(i)}e^{-i\xi_i{\sigma}_y^{(i)}/2}\ket{\Phi^{(i)}}=\dfrac{\eta_i}{\alpha n_i}\bra{\Psi^{(i)}}{\sigma}_x^{(i)}\ket{\Psi^{(i)}}.
\end{equation}
Applying the unitary $\ketbra{2^{(i)}}{2^{(i)}}+e^{-i\xi_i{\sigma}_y^{(i)}/2}$ to the state $r^{(i)}$, it becomes:
\begin{equation}
R'^{(i)}=n_i\ketbra{\Psi'^{(i)}}{\Psi'^{(i)}}+\left(1-n_i\right)\ketbra{2^{(i)}}{2^{(i)}},
\end{equation} 
 where $\ket{\Psi'^{(i)}}=e^{-i\xi_i{\sigma}_y^{(i)}/2}\ket{\Phi^{(i)}}$. Since $\commut{{\sigma}_y^{(i)}}{e^{-i\xi_i{\sigma}_y^{(i)}/2}}=0$, we have:
\begin{equation}
\bra{\Psi'^{(i)}}{\sigma}_y^{(i)}\ket{\Psi'^{(i)}}=\bra{\Phi^{(i)}}{\sigma}_y^{(i)}\ket{\Phi^{(i)}}=0.
\end{equation}
And because the state $\ket{\Psi'^{(i)}}$ is pure we have:
\[
\bra{\Psi'^{(i)}}{\sigma}_x^{(i)}\ket{\Psi'^{(i)}}^2+\bra{\Psi'^{(i)}}{\sigma}_y^{(i)}\ket{\Psi'^{(i)}}^2+\bra{\Psi'^{(i)}}{\sigma}_z^{(i)}\ket{\Psi'^{(i)}}^2=1
\]
i.e.,
\begin{equation}
\bra{\Psi'^{(i)}}{\sigma}_z^{(i)}\ket{\Psi'^{(i)}}^2=1-\dfrac{\eta_i^2}{\alpha^2 n_i^2}\bra{\Psi^{(i)}}{\sigma}_x^{(i)}\ket{\Psi^{(i)}}^2
\end{equation}

Finally we are in position to apply the Heisenberg uncertainty principle for the operators $S_x,S_y,S_z$ in the state $R'=\bigotimes R'^{(i)}$:
\begin{equation}
\label{detail}
\abs{\expect{S_x}{R'}}^2\leq \left(\Delta S_y\right)^2 \left(\Delta S_z\right)^2.
\end{equation}
For product states $R'=\bigotimes_{i=1}^N R'^{(i)}$, we have:
\begin{align}
\left(\Delta S_y\right)^2&=\sum_{i=1}^N\left(\Delta {\sigma}_y^{(i)}\right)^2=\sum_{i=1}^N\expect{\left({\sigma}_y^{(i)}\right)^2}{R'^{(i)}}-\expect{{\sigma}_y^{(i)}}{R'^{(i)}}^2\nonumber\\
&=\sum_{i=1}^Nn_i-\sum_{i=1}^Nn_i^2\bra{\Psi'^{(i)}}{\sigma}_y^{(i)}\ket{\Psi'^{(i)}}^2\nonumber \\
&=\sum_{i=1}^Nn_i. 
\end{align}
The same calculation for $S_z$ will give:
\bea
\left(\Delta S_z\right)^2&=&\sum_{i=1}^Nn_i-\sum_{i=1}^Nn_i^2\bra{\Psi'^{(i)}}{\sigma}_z^{(i)}\ket{\Psi'^{(i)}}^2\nonumber\\
&=&\sum_{i=1}^Nn_i-n_i^2+\dfrac{\eta_i^2}{\alpha^2}\bra{\Psi^{(i)}}{\sigma}_x^{(i)}\ket{\Psi^{(i)}}^2.
\eea
We also have:
\begin{equation}
\expect{S_x}{R'}=\sum_{i=1}^N\dfrac{\eta_i}{\alpha}\bra{\Psi^{(i)}}{\sigma}_x^{(i)}\ket{\Psi^{(i)}},
\end{equation}
but from Eq.~\eqref{eq:Ri}, we have:
\begin{equation}
\bra{\Psi^{(i)}}{\sigma}_x^{(i)}\ket{\Psi^{(i)}}=\dfrac{\expect{A_x^{(i)}}{\rho}}{\eta_i}.
\end{equation}
Using all the above the inequality~\eqref{detail}, can be simplified to obtain the desired form:
\begin{equation}
\label{eventighterAk_app}
\expect{A_x}{\rho}^2\leq \sum_{i=1}^Nn_i\left(\sum_{i=1}^N\expect{A_x^{(i)}}{\rho}^2\right)+\alpha^2\sum_{i=1}^Nn_i\left(\sum_{i=1}^Nn_i-\sum_{i=1}^Nn_i^2\right),
\end{equation}
which is exactly the inequality Eq.~\eqref{eventighterAk}, since $\expect{\hat{N}}{}=\sum_{i=1}^Nn_i$.
\subsubsection{Inequality for $\expect{A_i}{\rho}^2+\expect{A_j}{\rho}^2$}
One might suggest adding the two inequalities~\eqref{eventighterAk}
for the quantities $\expect{A_i}{\rho}^2$ and
$\expect{A_j}{\rho}^2$. But we can derive a tigther  inequality by considering the following mapping of the form~\eqref{eq:Ri}:
\bea
R^{(i)}=&n_i&\left(\dfrac{{\sigma_0}^{(i)}}{2}+\dfrac{\expect{A'^{(i)}_x}{\rho}}{2\eta_i}{\sigma_x}^{(i)}
+\dfrac{\expect{A'^{(i)}_y}{\rho}}{2\eta_i}{\sigma_y}^{(i)}+\dfrac{\expect{A'^{(i)}_z}{\rho}}{2\eta_i}\sigma_z\right)\nonumber\\&+&\left(1-n_i\right)\ketbra{2^{(i)}}{2^{(i)}}
\eea
where we have chosen:
$\expect{A'^{(i)}_x}{\rho}=\sqrt{\expect{A^{(i)}_x}{\rho}^2+\expect{A^{(i)}_y}{\rho}^2}$,
$\expect{A'^{(i)}_y}{\rho}=0$ and $\expect{A'^{(i)}_z}{\rho}=\expect{A^{(i)}_z}{\rho}$. Then we apply the inequality~\eqref{eventighterAk} for $\expect{A'^{(i)}_x}{\rho}$ and we get:
\bea
\expect{A'_x}{\rho}^2=&\left(\sum_{i=1}^N\sqrt{\expect{A^{(i)}_x}{\rho}^2+\expect{A^{(i)}_y}{\rho}^2}\right)^2&\nonumber\\
\leq &\expect{N}{\rho}\left(\sum_{i=1}^N\expect{A_x^{(i)}}{\rho}^2+\expect{A_y^{(i)}}{\rho}^2\right)&\nonumber\\&+\alpha^2\expect{N}{\rho}\left(\expect{N}{\rho}-\sum_{i=1}^N\expect{N^{(i)}}{\rho}^2\right).
\eea
Using Cauchy-Schwartz inequality \[ x_ix_j+y_iy_j\leq \sqrt{x_i^2+y_i^2}\sqrt{x_j^2+y_j^2}\] we obtain:
\begin{equation}
\left(\sum_{i=1}^N\expect{A^{(i)}_x}{\rho}\right)^2+\left(\sum_{i=1}^N\expect{A^{(i)}_y}{\rho}\right)^2\leq \left(\sum_{i=1}^N\sqrt{\expect{A^{(i)}_x}{\rho}^2+\expect{A^{(i)}_y}{\rho}^2}\right)^2
\end{equation}
and we finally get:
\bea
\expect{A_i}{\rho}^2+\expect{A_j}{\rho}^2&\leq \expect{N}{\rho}\left(\sum_{i=1}^N\expect{A_i^{(i)}}{\rho}^2+\expect{A_j^{(i)}}{\rho}^2\right)
\nonumber\\
&+\alpha^2\expect{N}{\rho}\left(\expect{N}{\rho}-\sum_{i=1}^N\expect{N^{(i)}}{\rho}^2\right).
\eea

Following the same steps, we can prove in general that:
\bea
\label{eq:Ak_general}
\sum_{k\in\mathcal{I}}\expect{A_k}{\rho}^2&\leq &\expect{N}{\rho}\sum_{i=1}^N\sum_{k\in\mathcal{I}}\expect{A_k^{(i)}}{\rho}^2
\nonumber\\
&+&\alpha^2\expect{N}{\rho}\left(\expect{N}{\rho}-\sum_{i=1}^N\expect{N^{(i)}}{\rho}^2\right)
\eea

Where $\mathcal{I}$ is any subset of $\lbrace 1,2,\cdots,M\rbrace$, and
\beq
\frac{\sum_{k=1}^{M} \expect{A^{(i)}_k}{\rho^{(i)}}^2}{\alpha^2}\leq \expect{n^{(i)}}{\rho^{(i)}}^2\leq 1
\eeq
\beq
\alpha^2 =
\text{sup}_{\rho^{(i)}} \left[\sum_{k=1}^{M} \expect{A^{(i)}_k}{\rho^{(i)}}^2\right].
\eeq
Notice that in the case of angular momentum operators $M=3$, as in the main text.
With inequality Eq.\eqref{eq:Ak_general}, we have all the ingredients
needed to derive inequality Eq.\eqref{eq:OurMainineq}.
\subsubsection{Proof of Eq.~\eqref{eq:OurMainineq} for product states}
Let $\mathcal{I}\subseteq \lbrace 1,\cdots,M\rbrace$ including the empty set $\phi$. We have the following equalities for product states:
\begin{eqnarray}
\expect{\tilde{A}_k^2}{}&\equiv &\expect{A_k^2}{}-\sum_{i}\expect{{A_k^{(i)}}^2}{}=\sum_{i\neq j}\expect{{A_k^{(i)}}{A_k^{(j)}}}{}
\nonumber\\
&=&\sum_{i\neq j}\expect{{A_k^{(i)}}}{}\expect{{A_k^{(j)}}}{}=\expect{A_k}{}^2-\sum_{i=1}^N\expect{A_k^{(i)}}{}^2 \label{Aktild}\\
\tilde{\Delta}^2A_k&\equiv&\expect{\tilde{A}_k^2}{}-\expect{A_k}{}^2=-\sum_{i=1}^N\expect{A_k^{(i)}}{}^2.\label{DeltaAktild}
\end{eqnarray}

From Eq.~\eqref{Aktild} and Eq.~\eqref{eq:Ak_general}, we get:
\bea
\label{tildinequality}
\sum_{k\in\mathcal{I}}\expect{\tilde{A}_k^2}{\rho}&\leq \left(\expect{N}{\rho}-1\right)\left(\sum_{i=1}^N\sum_{k\in\mathcal{I}}\expect{A_k^{(i)}}{\rho}^2\right)\nonumber\\
&+\alpha^2\expect{N}{\rho}\left(\expect{N}{\rho}-\sum_{i=1}^N\expect{N^{(i)}}{\rho}^2\right).
\eea
Now we have all the ingredients to derive the desired inequality. From~\eqref{tildinequality} and~\eqref{DeltaAktild} we get:  
\begin{widetext}
\begin{equation}
\label{beforeMain}
(\expect{\hat{N}}{\rho}-1)\sum_{k\notin \mathcal{I}}\tilde{\Delta}^2A_k-\sum_{k\in \mathcal{I}}\expect{\tilde{A}_k^2}{}\geq -\alpha^2\expect{{N}}{\rho}\left(\expect{N}{\rho}-\sum_{i=1}^N\expect{N^{(i)}}{\rho}^2\right)-\left(\expect{\hat{N}}{\rho}-1\right)\left(\sum_{i=1}^N\expect{A_x^{(i)}}{\rho}^2+\expect{A_y^{(i)}}{\rho}^2+\expect{A_z^{(i)}}{\rho}^2\right).
\end{equation}
\end{widetext}
The above inequality is hard to extend for mixed states, that's why we put it in a more convenient form and we simplify it further using $\expect{A_x^{(i)}}{\rho}^2+\expect{A_y^{(i)}}{\rho}^2+\expect{A_z^{(i)}}{\rho}^2\leq \alpha^2\expect{N^{(i)}}{\rho}^2$ to finally get: 

\begin{equation}
\label{Main}
\expect{\hat{N}}{\rho}\sum_{k\notin \mathcal{I}}\tilde{\Delta}^2A_k+\sum_{k\in \mathcal{I}}\tilde{\Delta}^2A_k-\sum_{k\in \mathcal{I}}\expect{\tilde{A}_k^2}{}\geq-\alpha^2\expect{{N}}{\rho}^2.
\end{equation}
\subsubsection{Proof of Eq.~\eqref{eq:OurMainineq} for  mixed separable states}
Let us consider the most general separable  state as a convex mixture of pure product states:
\begin{equation}
\rho=\sum_l\lambda_l\rho_l \quad: \lambda_l\geq 0,\, \sum_l\lambda_l=1,\, \rho_l^2=\rho_l
\end{equation}
 Then we have the following inequality :
 \bea
\sum_{k\notin \mathcal{I}}\tilde{\Delta}^2A_k+\alpha^2\expect{{N}}{\rho}\hspace{-0.7cm}&\underset{\text{\vspace{-0.5cm}concavity of variance}}{\geq}\hspace{-1cm}\sum_l\lambda_l\left(\sum_{k\notin \mathcal{I}}\tilde{\Delta}_l^2A_k+\alpha^2\expect{{N}}{\rho_l}\right)
\nonumber\\
&\underset{Eq.~\eqref{Main}}{\geq} \sum_l\lambda_l\sum_{k\in
  \mathcal{I}}\dfrac{\expect{{A_k}}{\rho_l}^2}{\expect{{N}}{\rho_l}} \label{eq:toconvex},
 \eea
 where we have used the definition of the modified moments
 $\tilde{\Delta}^2A_k=\expect{\tilde{A}_k^2}{}-\expect{A_k}{}^2$ for
 the right hand side. 
Next, we use the convexity of the function
$f\left(x,y\right)=\dfrac{x^2}{y}$ over
$\mathbb{R}\times\left(0,\infty\right]$. The convexity of $f(x,y)$ can
be shown by considering its Hessian matrix
 $H_{i,j}=\dfrac{\partial^2f\left(x_1,x_2\right)}{\partial x_i\partial x_j}$:
 \begin{equation}
 H=
\begin{pmatrix}
\dfrac{2}{y} & -\dfrac{2x}{y^2} \\ -\dfrac{2x}{y^2} & \dfrac{2x^2}{y^3}
\end{pmatrix} 
 \end{equation}
 which eigenvalues  are $\lbrace
 0,\dfrac{2(x^2+y^2)}{y^3}\rbrace$. Both eigenvalues being positive for any $\left(x,y\right)\in \mathbb{R}\times\left(0,\infty\right]$, we
 can conclude $f\left(x,y\right)=\dfrac{x^2}{y}$ is convex. From the convexity of $f\left(x,y\right)$, we obtain a lower bound of Eq.~\eqref{eq:toconvex}:
 \begin{equation}
\sum_l\lambda_l\sum_{k\in \mathcal{I}}\dfrac{\expect{{A_k}}{\rho_l}^2}{\expect{{N}}{\rho_l}}\geq \sum_{k\in \mathcal{I}}\dfrac{\expect{{A_k}}{\rho}^2}{\expect{{N}}{\rho}}=\dfrac{ \sum_{k\in \mathcal{I}}\expect{\tilde{A}_k^2}{\rho}-\sum_{k\in \mathcal{I}}\tilde{\Delta}^2A_k}{\expect{{N}}{\rho}}
 \end{equation}
 completing the proof of inequality Eq.~\eqref{eq:OurMainineq} for any separable state.

\section{Mapping to a qutrit Eq~\eqref{mappingtoqutrit}}\label{map}
Here we will consider only a special case of mappings to a qutrit, where the image is a mixed state of a spin-$1/2$ like state, in the subspace $\ket{0},\ket{1}$, and the state $\ket{2}$ as in  Eq~\eqref{mappingtoqutrit}.  
The starting point is the mapping that maps every spin-$j$ state to the following spin-$1/2$ state:
\begin{equation}
\mathcal{M}\left(\rho\right)=\dfrac{\openone}{2}+\dfrac{\expect{A_x}{\rho}}{2\eta}\sigma_x+\dfrac{\expect{A_y}{\rho}}{2\eta}\sigma_y+\dfrac{\expect{A_z}{\rho}}{2\eta}\sigma_z,
\end{equation}
where $\eta= \sqrt{\expect{A_x}{\rho}^2+\expect{A_y}{\rho}^2+\expect{A_z}{\rho}^2} $. The above mapping is a completely positive mapping that can be written as~\citep{Saideh2015}
\begin{equation}
\mathcal{M}\left(\rho\right)=\text{Tr}_{\mathcal{H}_D}\left[U\rho\otimes \ketbra{0}{0}U^{\dagger}\right],
\end{equation}
where $\ket{0}$ is a reference state in the qubit subspace and  $U$ is an isometry that can always be written as:
\begin{equation}
U : \mathcal{H}_{2s+1}\otimes\mathcal{H}_2\rightarrow \mathcal{H}_{D}\otimes\mathcal{H}_2 : U=\sum_{i=0}^4\mathcal{A}_i\otimes \sigma_i\,;\,\sigma_0=\openone,
\end{equation}
where the dimension of $\mathcal{H}_{D}$ satisfies $D\geq 2s+1$.
Since the three Gell-Mann matrices $\Lambda_{1,2}^s$, $\Lambda_{1,2}^a$ and $\Lambda_1$~\cite{gellman} are the Pauli operators in the subspace $\lbrace\ket{0},\ket{1}\rbrace$ of the qutrit, we can define the following mapping:
 \begin{equation}
\mathcal{M}\left(\rho\right)=\text{Tr}_{\mathcal{H}_D}\left[\mathcal{U}\rho\otimes \left(\beta\ketbra{0}{0}+\left(1-\beta\right)\ketbra{2}{2}\right)\mathcal{U}^{\dagger}\right],
\end{equation}
where
\begin{equation}
\mathcal{U} : \mathcal{H}_{2s+1}\otimes\mathcal{H}_3\rightarrow \mathcal{H}_{D}\otimes\mathcal{H}_3 : \mathcal{U}=\sum_{i=0}^4\mathcal{A}_i\otimes \sigma_i +\mathcal{I}\otimes\ketbra{2}{2},
\end{equation}
\begin{equation}
\sigma_0=\ketbra{0}{0}+\ketbra{1}{1}\,,\, \sigma_1=\Lambda_{1,2}^s\,,\, \sigma_2=\Lambda_{1,2}^a\,,\, \sigma_3=\Lambda_{1}\nonumber
\end{equation}
\begin{equation}
\mathcal{I} : \mathcal{H}_{2s+1}\rightarrow \mathcal{H}_{D}:\, \mathcal{I}^{\dagger}\mathcal{I}=\openone_{2s+1}
\end{equation}
 an arbitrary isometry from $\mathcal{H}_{2s+1}$ to $\mathcal{H}_{D}$, and $0<\beta<1$ is some positive number which, in the mapping of
interest~\eqref{mappingtoqutrit}, was set to be
$\beta_i=n_i=\expect{N^{(i)}}{\rho}$. 
One can easily verify that $\mathcal{U}$ is an isometry, i.e.,
$\mathcal{U}^{\dagger}\mathcal{U}=\openone_{2s+1}\otimes\openone_{3}$ 
and that the resulting mapping can be written as:
\begin{align}
\mathcal{M}\left(\rho\right)&=\beta\left(\dfrac{\sigma_0}{2}+\dfrac{\expect{A_x}{\rho}}{2\eta}\sigma_x+\dfrac{\expect{A_y}{\rho}}{2\eta}\sigma_y
+\dfrac{\expect{A_z}{\rho}}{2\eta}\sigma_z\right) \nonumber \\
&+\left(1-\beta\right)\ketbra{2}{2}.
\end{align}
Hence we have proven that the mapping~\eqref{mappingtoqutrit} is 
completely positive.

\end{document}